\documentclass[prl,twocolumn,floatfix,nofootinbib,amsmath,amssymb,floatfix]{revtex4} 
\usepackage{graphicx,color,dcolumn,booktabs,bm,mathrsfs}
\usepackage{longtable,lscape}
\usepackage{txfonts}
\usepackage{overpic}
\usepackage{amssymb}
\usepackage{indentfirst}
\usepackage{feynmf}   
\usepackage{slashed}  
\usepackage{cases}
\usepackage{color}
\usepackage{multirow}
\usepackage{epstopdf}
\usepackage{CJK}

\usepackage[colorlinks,
            citecolor=blue,
            anchorcolor=red,
            menucolor=red,
            linkcolor=red,
            filecolor=red,
            runcolor=red,
            urlcolor=blue,
            frenchlinks=red]{hyperref}
\usepackage{subfigure}
\usepackage{xcolor}

\newcommand{\be}{\begin{equation}} 
\newcommand{\ee}{\end{equation}}
\newcommand{\bea}{\begin{eqnarray}} 
\newcommand{\eea}{\end{eqnarray}}

\begin{document}

\title{Searching for the two poles of the $\Xi(1820)$ in the $\psi(3686) \to \bar{\Xi}^+ \bar{K}^0 \Sigma^{*-}(\pi^- \Lambda)$ decay}


\author{Man-Yu Duan$^{1,2}$}\email{duanmy@seu.edu.cn}
\author{Jing Song$^{1,3}$}\email{Song-Jing@buaa.edu.cn}
\author{Wei-Hong Liang$^{4,5}$}\email{liangwh@gxnu.edu.cn}
\author{E. Oset$^{1,4}$}\email{oset@ific.uv.es}

\affiliation{$^1$Departamento de Física Teórica and IFIC, Centro Mixto Universidad de Valencia-CSIC Institutos de Investigación de Paterna, 46071 Valencia, Spain\\
$^2$School of Physics, Southeast University, Nanjing 210094, China\\
$^3$School of Physics, Beihang University, Beijing, 102206, China\\
$^4$Department of Physics, Guangxi Normal University, Guilin 541004, China\\ 
$^5$Guangxi Key Laboratory of Nuclear Physics and Technology, Guangxi Normal University, Guilin 541004, China}

\begin{abstract}

We propose the reaction $\psi(3686) \to \bar{\Xi}^+ \bar{K}^0 \Sigma^{*-}$, with the $\Sigma^{*-}$ decaying to $\pi^- \Lambda$ in order to show evidence for the existence of two $\Xi(1820)$ states, one around $1824$ MeV and narrow, and another one around $1875$ MeV and wide. The phase space for $\bar{K}^0 \Sigma^{*-}$ production reduces the effect of the lower mass resonance, magnifying the effect of the higher mass resonance that shows clearly over the phase space. The estimated rate  of the production is bigger than the one of the $\psi(3686) \to \bar{\Xi}^+ K^- \Lambda$ reaction, where a clear peak for $\Xi(1820)$ was observed by the BESIII collaboration, what makes the Beijing facility ideal to carry out the reaction proposed.

\end{abstract}

\maketitle

The recent BESIII experiment on the $\psi(3686)$ decay into $K^- \Lambda \bar{\Xi}^+$ \cite{BESIII:2023mlv}, where the $K^- \Lambda$ spectrum showed a peak in the $\Xi(1820)$ region with an unexpected width of about $73$ MeV, spurred theoretical work reclaiming the existence of two $\Xi(1820)$ states. This large width contrasts with the PDG average of $24 \pm 5$ MeV \cite{Workman:2022ynf}, and the large discrepancy reopened the issue of the two $\Xi(1820)$ states predicted in Ref.\cite{Sarkar:2004jh} within the chiral unitary approach. This theoretical framework, applied to the present case, studied the interaction of pseudoscalar mesons with the $J^P = \frac{3}{2}^+$ baryons of the $\Delta$ decuplet and found many resonant states dynamically generated from that interaction, which matched existing states in the $N$, $\Delta$, $\Lambda$, $\Sigma$, $\Xi$ and $\Omega$ sectors with $J^P = \frac{3}{2}^-$ \cite{Kolomeitsev:2003kt,Sarkar:2004jh}. One of the predictions of Ref.\cite{Sarkar:2004jh} was an $\Omega$ state originated from the $K^-\Xi(1530)$ and $\eta\Omega$ interaction, which was later identified with the recently found $\Omega(2012)$ state by the Belle collaboration \cite{Belle:2018mqs}. A discussion followed on the nature of this state (see Ref. \cite{Ikeno:2022jpe} for the latest update), which finally led the Belle collaboration to conclude that the experimental information supported the molecular nature of this resonance \cite{Belle:2022mrg}.

Coming back to the $\Xi(1820)$ states, a state at around $1820$ MeV was found in Refs.\cite{Sarkar:2004jh,Kolomeitsev:2003kt}, another $\Xi$ state was found around $2100$ MeV, and in Ref.\cite{Sarkar:2004jh} a wider pole was found in the complex plane in the $1800-1900$ MeV region.

The issue of two poles associated to some known resonances has caught up after the two poles predicted for the $\Lambda(1405)$ \cite{Oller:2000fj,Jido:2003cb} were officially admitted in the PDG. Examples of this can be found in Ref.\cite{Geng:2006yb} with two states for the $K_1(1270)$, the two poles of the $D^*(2400)$ \cite{Albaladejo:2016lbb}, and the two $Y(4260)$ states of BESIII \cite{BESIII:2020bgb}. Recently a paper \cite{Xie:2023cej} shows that the phenomenon of duplication of states is tied to the structure of the Weinberg-Tomozawa interaction. A review on the issue of the two poles of some resonances can be seen in Ref.\cite{Meissner:2020khl}.

From this perspective, the issue of  the two $\Xi(1820)$ poles was retaken in the work of Ref.\cite{Molina:2023uko}, looking at the interaction of the $\Sigma^* \bar{K}$, $\Xi^* \pi$, $\Xi^* \eta$ and $\Omega K$ coupled channels, and by means of it a good description of the BESIII data was obtained, with two poles at $1824-31 i$ MeV and $1875-130 i$ MeV. Work continued in Ref.\cite{Liang:2024fsv} proposing the $\Omega_c \to \pi^+ (\pi^0, \eta) \pi \Xi^*$ reactions, by means of which an interference pattern between the two resonances was found in the $\pi(\eta) \Xi^*$ invariant mass distributions that could shed extra information leading to the identification of the two poles. In these reactions the two states interfered in such a way that a dip is seen around $1850$ MeV, in a pattern that recalls the interference between the $f_0(500)$ and $f_0(980)$ in $I=0$ $S$-wave $\pi \pi$ scattering, leading to a dip in the $\pi \pi$ cross section around the $f_0(980)$ region \cite{Protopopescu:1973sh}.

In the present work we propose a reaction, which is a continuation of the BESIII experiment with $\psi(3686)$ decay, but in a different channel. In Ref.\cite{BESIII:2023mlv} the $K^- \Lambda$ mass distribution in $\psi(3686) \to \bar{\Xi}^+ K^- \Lambda$ was investigated, with the $K^- \Lambda$ mass far below the $1820$ MeV region, and mostly the narrow resonance at $1824$ MeV showed up, with the wider resonance providing strength in the higher energy region. In the work of Ref.\cite{Liang:2024fsv} the weight of the resonances is different and gives rise to an interference pattern.

The reaction we propose is meant to show in a clear way the higher mass resonance. For this we propose to look at the $\bar{K} \Sigma^*$ final state in the $\psi(3686) \to \bar{\Xi}^+ \bar{K}^0 \Sigma^{*-}$ reaction, which has a threshold at about $1880$ MeV, although we can attain smaller energies through the tail of the $\Sigma^*(1385)$ resonance, where the strength of the $\Xi(1820)$ state at $1824$ MeV has been drastically reduced, then giving more room to the wide $\Xi(1820)$ state at $1875$ MeV. The reaction proposed is thus: $\psi(3686) \to \bar{\Xi}^+ \bar{K}^0 \Sigma^{*-} \to \bar{\Xi}^+ \bar{K}^0 \pi^- \Lambda$, where the $\bar{K}^0 \pi^- \Lambda$ mass can go down around $1750$ MeV, but the highest strength appears around $1950$ MeV.

\begin{figure*}[t]
\centering
\includegraphics[scale=0.75]{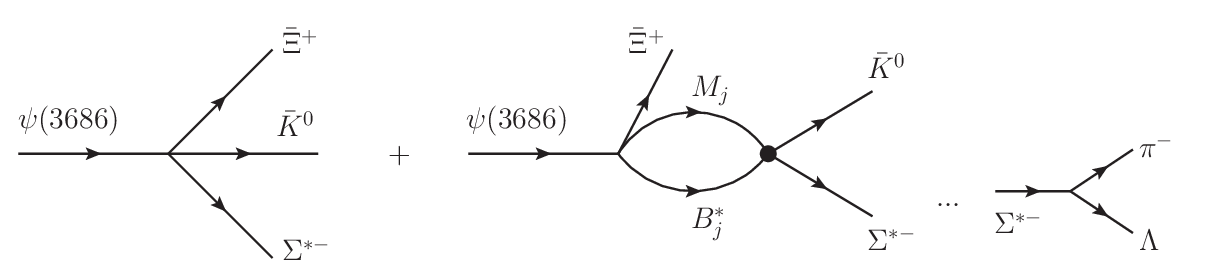} \put(-400,-5){(a)} \put(-200,-5){(b)} \put(-60,-5){(c)}
\caption{Mechanisms for $\bar{K}^0 \Sigma^{*-}$ production: (a) tree level; (b) rescattering of the meson-baryon coupled channels; (c) decay of $\Sigma^{*-}$ into the stable $\pi^- \Lambda$ particles.}
\label{fig:mech}
\end{figure*}

We recall that the two $\Xi(1820)$ states were obtained as poles of the $T$ matrix from the interaction of the coupled channels, $ \bar{K}^0 \Sigma^{*-}$, $ K^- \Sigma^{*0}$, $ \pi^0 \Xi^{*-}$, $ \eta \Xi^{*-}$, $ \pi^- \Xi^{*0}$, $ K^0 \Omega^-$ (for the $\Xi(1820)^-$) \cite{Molina:2023uko}, via the Bethe-Salpeter equation
\begin{equation}\label{eq:BS}
  T=[1-VG]^{-1} \, V,
\end{equation}
where
\begin{equation}\label{eq:Vij}
V_{ij}=-\dfrac{1}{4f^2} C_{ij} (k^0+k^{\prime \,0});\, f=1.28f_{\pi},\, f_{\pi}=93\, \mathrm{MeV}, 
\end{equation}
with $k^0$, $k^{\prime \,0}$ the energies of the initial and final pseudoscalar mesons in the meson-baryon rest frame, and $C_{ij}$ the coefficients given in Tables A.4.2 and A.4.3 of Ref.\cite{Sarkar:2004jh}. $G$ in Eq.~\eqref{eq:BS} is the diagonal meson-baryon loop function regularized with a cutoff $q_{\mathrm{max}}$, as in Ref.\cite{Oset:1997it}. We take $q_{\mathrm{max}}=830 \, \mathrm{MeV}$, same as in Ref.\cite{Molina:2023uko}. The mechanism for $\bar{K}^0 \Sigma^{*-}$ production is depicted in Fig.~\ref{fig:mech}.

Analytically, the transition matrix of Fig.~\ref{fig:mech} corresponds to
\begin{equation}\label{eq:t}
t=C \, \langle B^* \,| \, (\vec{S}^+ \times \vec{p}_{\bar{\Xi}^+}) \cdot \vec{\epsilon} \, | \, \Xi^- \rangle \, t',
\end{equation}
where $C$ is a global normalization constant, $B^*$ is the baryon of the $\frac{3}{2}^+$ multiplet, $\vec{\epsilon}$ the vector polarization of the $\psi(3686)$ and $\vec{p}_{\bar{\Xi}^+}$ the momentum of the $\bar{\Xi}^+$ in the rest frame of the $\psi(3686)$. The operator $\vec{S}^+$ is the spin transition operator from spin $\frac{1}{2}$ to $\frac{3}{2}$ with the property,
\begin{equation}\label{eq:s}
\sum_M S_i \,|\, M \rangle \,\langle M \,|\, S_j^{\dagger} = \frac{2}{3}\, \delta_{ij} - \frac{i}{3}\, \epsilon_{ijl}\, \sigma_l \, .
\end{equation}
The matrix $t'$ is then given by
\begin{equation}\label{eq:tp}
t'=W_{ \bar{K}^0 \Sigma^{*-}}+\sum_j W_j \, G_j \, t_{j,\bar{K}^0 \Sigma^{*-} }\,,
\end{equation}
where $t_{j, \bar{K}^0 \Sigma^{*-}}$ are matrix elements stemming from Eq.~\eqref{eq:BS}, and $W_j$, with $j$ corresponding to any of the six coupled channels, are the weights for the first step production $\psi(3686) \to \bar{\Xi}^+ M_j B^*_j$, which are calculated as follows: the $\psi(3686)$, being a $c\bar{c}$ state, is a singlet of $SU(3)$ (for $u$, $d$, $s$ quarks). As a consequence, up to a global normalization accounted for by the factor $C$ in Eq.~\eqref{eq:t}, these coefficients are the $SU(3)$ Clebsch-Gordan coefficients of $8 \otimes 10 \to 8$, choosing for the $8 \otimes 10$ the $MB^*$ states of the coupled channels, and for the final $8$ multiplet the state $\Xi^-$. The coefficients $W_j$ are given in Table~\ref{tab:Wj}.

\renewcommand\arraystretch{2}
\begin{table}[b]
 \begin{center}
\caption{$W_j$ Clebsch-Gordan coefficients for the different coupled channels.}
\label{tab:Wj}
\setlength{\tabcolsep}{1.5mm}{
\begin{tabular}{c c c c c c c}
 \toprule[1 pt]
Channels   & $ \bar{K}^0 \Sigma^{*-}$ & $ K^- \Sigma^{*0}$ & $ \pi^0 \Xi^{*-}$ & $ \eta \Xi^{*-}$ & $ \pi^- \Xi^{*0}$ & $ K^0 \Omega^-$  \\
 \midrule[1 pt]
$W_j$   & $-\sqrt{\frac{2}{15}}$  &  $-\sqrt{\frac{1}{15}}$  &  $\sqrt{\frac{1}{15}}$  &  $-\sqrt{\frac{1}{5}}$ & $-\sqrt{\frac{2}{15}}$ & $\sqrt{\frac{2}{5}}$  \\
 \bottomrule[1 pt]
\end{tabular}}
  \end{center}
\end{table}

The coefficients $W_j$ already account for the isospin phase convention $|K^-\rangle=-|\frac{1}{2},-\frac{1}{2}\rangle$, $|\pi^+\rangle=-|1,1\rangle$, consistent with the convention used in Refs.\cite{Sarkar:2004jh,Molina:2023uko} and the Clebsch-Gordan coefficients of Ref.\cite{McNamee:1964xq} used here. The intrinsic phase of the $SU(3)$ Clebsch-Gordan coefficients for the mesons can be obtained from $8 \otimes 8 \to 1$ demanding to get the symmetrical combination $K^+K^- + K^0\bar{K}^0 + \pi^+\pi^- + \pi^0\pi^0 + \pi^-\pi^+ + \eta\eta + K^-K^+$. The phase convention for the decuplet baryons is obtained in a similar way, demanding that  $10 \otimes 10 \to 1$ gives a symmetrical combination of all states and their antiparticles, which results into the isospin multiplets ($\Delta^{++}$, $\Delta^{+}$, $\Delta^{0}$, $\Delta^{-}$), ($\Sigma^{*+}$, $\Sigma^{*0}$, $\Sigma^{*-}$), ($\Xi^0$, $\Xi^-$), as assumed in Ref.\cite{Sarkar:2004jh}, but the antiparticles of $\Delta^{++}$, $\Delta^{0}$, $\Sigma^{*0}$, $\Xi^{*0}$ carry a negative phase.

The mass distribution for the decay $\psi(3686) \to \bar{\Xi}^+ \bar{K}^0 \Sigma^{*-}$ is given by
\begin{equation}\label{eq:gam1}
\frac{d\Gamma}{dM_{\mathrm{inv}}(\bar{K}^0 \Sigma^{*-})}=\frac{1}{(2\pi)^3} \frac{1}{4M_{\psi}^2} p_{\bar{\Xi}^+} \tilde{p}_{\bar{K}^0} \bar{\sum} \sum |t|^2 2 M_{\bar{\Xi}^+} 2 M_{\Sigma^{*-}} \,,
\end{equation}
with
\begin{equation}\label{eq:ssum}
 \bar{\sum} \sum |t|^2=\frac{8}{9}C^2 |t'|^2 p_{\bar{\Xi}^+}^2\,,
 \end{equation}
where $\tilde{p}_{\bar{K}^0}$ is the momentum of the $\bar{K}^0$ in the $\bar{K}^0 \Sigma^{*-}$ rest frame. We can gather some constant factors together and write
\begin{eqnarray}\label{eq:gam2}
\frac{d\Gamma}{dM_{\mathrm{inv}}(\bar{K}^0 \Sigma^{*-})}&=&\frac{1}{(2\pi)^3} \frac{1}{4M_{\psi}^2} p_{\bar{\Xi}^+} \tilde{p}_{\bar{K}^0} |t'|^2 \frac{C'}{M^2_{\psi}} p_{\bar{\Xi}^+}^2 \nonumber\\
 &=& \frac{1}{(2\pi)^3} \frac{C'}{4M_{\psi}^4} p^3_{\bar{\Xi}^+} \tilde{p}_{\bar{K}^0} |t'|^2 \,,
\end{eqnarray}
with $C'$ a dimensionless constant.

We can go one step forward to consider the mass distribution of the $\Sigma^{*-}$ and the branching ratio for $\Sigma^{*-} \to \pi^- \Lambda$ decay and write
\begin{widetext}
\begin{equation}\label{eq:gamCal}
\frac{d\Gamma}{dM_{\mathrm{inv}}(\bar{K}^0 \Sigma^{*-}) dM_{\mathrm{inv}}(\Sigma^{*-})}=-\frac{1}{\pi}\mathrm{Im}\frac{\frac{\Gamma_{\pi^-\Lambda}}{\Gamma_{\Sigma^{*-}}}}{M_{\mathrm{inv}}(\Sigma^{*-})-M_{\Sigma^{*-}}+i\frac{\Gamma_{\Sigma^{*-}}\left(M_{\mathrm{inv}}(\Sigma^{*-})\right)}{2}} \cdot \frac{1}{(2\pi)^3} \frac{C'}{4M_{\psi}^4} p^3_{\bar{\Xi}^+} \tilde{p}_{\bar{K}^0} |t'|^2\,,
\end{equation}
\end{widetext}
with $\tilde{p}_{\bar{K}^0}$ given now by
\begin{equation}\label{eq:pknew}
\tilde{p}_{\bar{K}^0}=\frac{\lambda^{1/2}\left(M^2_{\mathrm{inv}}(\bar{K}^0 \Sigma^{*-}), m^2_{\bar{K}^0}, M^2_{\mathrm{inv}}(\Sigma^{*-})\right)}{2M_{\mathrm{inv}}(\bar{K}^0 \Sigma^{*-})} \, .
\end{equation}
In addition, we take the width of the $\Sigma^{*-}$ energy dependent as
\begin{equation}\label{eq:ongam}
\Gamma_{\Sigma^{*-}}\left(M_{\mathrm{inv}}(\Sigma^{*-})\right)=\Gamma_{\mathrm{on}} \frac{M_{\Sigma^{*-}}}{M_{\mathrm{inv}}(\Sigma^{*-})} \left( \frac{\tilde{p}_{\pi}}{\tilde{p}_{\pi,\mathrm{on}}} \right)^3\,,
\end{equation}
with $\Gamma_{\mathrm{on}}$ the width of $\Sigma^{*-}$, $\tilde{p}_{\pi}$ the $\pi^-$ momentum in the decay of a $\Sigma^{*-}$ of invariant mass $M_{\mathrm{inv}}(\Sigma^{*-})$ into $\pi^- \Lambda$ and $\tilde{p}_{\pi,on}$ the same momentum for the nominal mass of the $\Sigma^{*-}$. This formula assumes the energy dependence to be due totally to the $\Sigma^{*-} \to \pi^- \Lambda$ decay, a good approximation when $\Gamma_{\pi^-\Lambda}/ \Gamma_{\Sigma^{*-}}=87\%$.

\begin{figure}[h]
\centering
\includegraphics{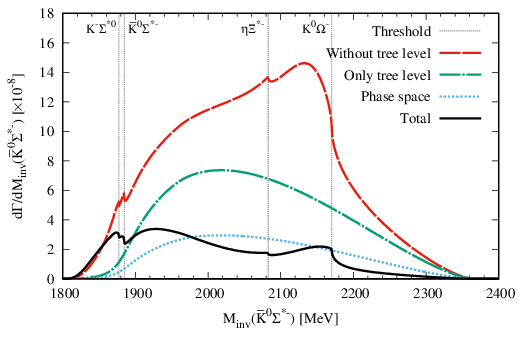}
\caption{$d\Gamma/dM_{\mathrm{inv}}(\bar{K}^0 \Sigma^{*-})$ with different options.The thresholds of the different channels are shown by gray vertical lines. Red dashed line: results without the tree level. Green dot-dashed line: results with only the tree level. Blue dotted line: phase space (tree level normalized to the area of the full results). Black continuous line: full results.}
\label{fig:resu}
\end{figure}

In Fig.~\ref{fig:resu} we show the results that we obtain for $d\Gamma/dM_{\mathrm{inv}}(\bar{K}^0 \Sigma^{*-})$ by integrating Eq.~\eqref{eq:gamCal} over the mass distribution of the $\Sigma^{*-}$, $dM_{\mathrm{inv}}(\Sigma^{*-})$. We show there several curves. The upper curve results from ignoring the tree level (term with $W_{\bar{K}^0 \Sigma^{*-}}$ in $t'$ of Eq.~\eqref{eq:tp}). The vertical lines correspond to the threshold of the channels $K^- \Sigma^{*0}$, $\bar{K}^0 \Sigma^{*-}$, $\eta \Xi^{*-}$, $K^0 \Omega^-$. We can see that thanks to the mass distribution of the $\Sigma^{*-}$ we can go below the nominal $K^- \Sigma^{*0}$ threshold. This allows the contribution of the low mass $\Xi(1820)$, but suppressed by the phase space. This feature is important because then the bulk of the strength of the mass distribution corresponds to the higher mass $\Xi(1820)$ resonance. This is the main purpose of this reaction, which is showing evidence for the higher mass resonance, since in other reactions the lower mass $\Xi(1820)$ plays a dominant role. In Fig.~\ref{fig:resu} we also show the results obtained with only tree level, which is sizeable, and with the black continuous line we show the full results using all terms in $t'$. It is compared with the tree level normalized to the same area, which is what would correspond to the phase space for the reaction. What we observe is that the actual mass distribution differs appreciably from phase space. We can see that below $1880$ MeV there is interference of the tree level and the two resonances, but the excess of strength above $1900$ MeV is due to the wide $\Xi(1820)$ resonance of higher energy. In addition we also observe strength in the region of $2100-2200$ MeV due to a resonance obtained in this region in Refs.\cite{Sarkar:2004jh,Kolomeitsev:2003kt}, which should also show up in this reaction. Yet, our main concern here is to show that the proposed reaction is particularly suited to show the effect of the second $\Xi(1820)$ resonance predicted theoretically.

We have not calculated the absolute value of the strength of the mass distribution. However, since in BESIII the $\psi(3686) \to \bar{\Xi}^+ K^- \Lambda$ reaction \cite{BESIII:2023mlv} was observed with good statistics, and, as shown in Ref.\cite{Molina:2023uko}, this reaction involves the same mechanism of $\psi(3686) \to \bar{\Xi}^+ M B^*$ of Fig.~\ref{fig:mech}, plus the extra step of $M B^* \to K^- \Lambda$, it looks clear that the strength of the mass distribution of the reaction proposed here should be even bigger than the one observed in the BESIII experiment. With this perspective, we can only encourage the BESIII collaboration to perform this experiment that should show clear evidence of the existence of the two $\Xi(1820)$ resonances.

{\it Acknowledgments.} -- This work is partly supported by the National Natural Science Foundation of China (NSFC) under Grants No. 12365019 and No. 11975083, and by the Central Government Guidance Funds for Local Scientific and Technological Development, China(No. Guike ZY22096024). This work is also partly supported by the Spanish Ministerio de Economia y Competitividad (MINECO) and European FEDER funds under Contracts No. FIS2017-84038-C2-1-P B, PID2020-112777GB-I00, and by Generalitat Valenciana under contract PROMETEO/2020/023. This project has received funding from the European Union Horizon 2020 research and innovation programme under the program H2020-INFRAIA-2018-1, grant agreement No. 824093 of the STRONG-2020 project. This work is also partly supported by the National Natural Science Foundation of
China under Grants No. 12247108 and the China Postdoctoral Science Foundation under Grant No. 2022M720360 and  No. 2022M720359.


\end{document}